\documentclass[onecolumn]{revtex4}

\usepackage{graphicx}
\usepackage{dcolumn}
\usepackage{color}
\usepackage{bm}
\usepackage{amsmath}
\usepackage{amsfonts}
\usepackage{amssymb}

\input epsf

\begin{document}

\title{\bf Anomalous effective action, Noether current, Virasoro algebra and Horizon entropy}

\author{\bf Bibhas Ranjan Majhi\footnote{bibhas.majhi@mail.huji.ac.il, bibhas@iucaa.ernet.in}}
\affiliation{IUCAA, Post Bag 4, Ganeshkhind,\\
Pune University Campus, Pune 411 007, India
\vskip 3mm
and 
\vskip 3mm
Racah Institute of Physics, Hebrew University of Jerusalem,\\
Givat Ram, Jerusalem 91904, Israel}

\author{\bf Sumanta Chakraborty \footnote{sumanta@iucaa.ernet.in}}
\affiliation{IUCAA, Post Bag 4, Ganeshkhind,\\
Pune University Campus, Pune 411 007, India}


\begin{abstract}
Several investigations show that in a very small length scale there exists corrections to the entropy
of black hole horizon. Due to fluctuations of the background metric and the external fields
the action incorporates corrections. In the low energy regime, the one loop effective action in
four dimensions leads to trace anomaly.
We start from the Noether current corresponding to the Einstein-Hilbert plus the one loop effective action
to calculate the charge for the diffeomorphisms which preserve the 
Killing horizon structure. Then a bracket
among the charges is calculated. We show that the Fourier 
modes of the bracket is exactly similar to Virasoro
algebra. Then using Cardy formula the entropy is evaluated. 
Finally, the explicit terms of the entropy expression is calculated 
for a classical background. It turns out that the usual expression
for entropy; i.e. the Bekenstein-Hawking form, is not modified.
\end{abstract}
\maketitle


\section{Introduction and motivation}

One of the striking features of general relativity is its deep
connection with the laws of thermodynamics. A ``marriage''
between general theory of relativity (GR) and quantum mechanics
shows that black holes behave as thermodynamic objects with a
Hawking temperature $T_H=\kappa /2\pi$ and intrinsic entropy given
by Bekenstein-Hawking area law $S=A_h/4G$, where $\kappa$ is the
surface gravity and $A_h$ is the area of the horizon
\cite{Bekenstein} (for recent reviews see \cite{padrev}). It
should be noted that the interpretation of classical black hole
laws with that of black hole thermodynamics become possible due to
taking into account the quantum mechanical effects, leading to
Hawking radiation. Also we could readily obtain the same area
dependence of entropy and that leads to a kind of universality to
the above result.

Although the thermodynamic interpretations were first
observed in black hole solutions, such features are much more
general. It has been observed that the accelerated observer on a
Minkowski spacetime also associates temperature and entropy on the
Rindler horizon. This has been originated from the Unruh effect
\cite{Unruh}. This leads us to think that the static observer
(black hole) and the Rindler observer (accelerated frame) are in
the same footings. The reality is much more general. In a local
region one can always has a null surface which is not a solution
of the Einstein's equation and on that a entropy functional can be
associated such that evaluation of it on the null surface leads to
the entropy expression. Moreover, the extremization of it leads to
the Einstein's equations. This is very important in the context of
{\it emergent} paradigm of gravity \cite{pad11}. Such a local
concept implies that the notion of entropy is much more
fundamental and it must be observer dependent quantity. Hence any
analysis leading to it must have an observer dependent {\it
off-shell} description.

For more general class of theories including higher curvature terms in the action the Bekenstein-Hawking
area law no longer applies.
However, assuming the Zeroth law, the first law can be derived in these class
of covariant actions \cite{Wald}. Here the Noether charge corresponding to the diffeomorphism symmetry,
calculated over the spatial cross-sections of the horizon,
plays the role of
 entropy. This mode of thinking has two important aspects,
which are as follows:\\
$\bullet$ These results implies that the connection between gravity and thermodynamics is valid well
beyond Einstein theory of gravity. It only requires some general argument, like covariance and
principle of equivalence 
{\footnote{In some modified gravity theories without the equivalence 
principle one can construct black hole solutions, first laws, and thermal radiation \cite{Matt}}}.\\
$\bullet$ It is also possible to attribute an observer dependent entropy to any null surfaces
(need not be event horizons).

So far the results are well valid in the classical regime. In a
very small length scale, like the Plank length, the quantum
effects has to be incorporated, particularly since the
fluctuations of the metric and the external fields can not be
ignored. Then the classical results have to be modified. One
way to include these effects is to find the effective action
corresponding to the fluctuations. It modifies the partition
function and hence the entropy \cite{Birrell}. It turns out that
in GR the leading order correction to entropy is logarithmic of
the horizon area \cite{Davies77} and the coefficient of it is
related to the trace anomaly. Similar results has been achieved in
several methods (see for example \cite{Page}, \cite{Kaul},
\cite{Banerjee}) but the value of the coefficient depends on the
particular model for calculating the entropy. On the other hand, 
the entanglement of quantum fields between inside and outside of 
horizon approach leads to power law corrections \cite{Das:2007mj}. 
Hence it is evident that existence of corrections is universal and 
one would expect identical situation in a method of computing entropy.

One of the interesting ways to understand this universality is 
by Noether charge prescription. This will be done for the anomalous effective
action due to the fluctuations of the quantum fields which leads to the trace anomaly in four
dimensions. The one loop effective action can be determined by the effective field theory technique.
In the low energy limit, if one breaks the conformal symmetry the resulting theory leads to trace
anomaly \cite{Buchbinder} -- \cite{Mottola}.
In \cite{Aros}, an attempt has been done by evaluating the Noether charge corresponding
to the GR action plus this one loop effective action for a 
Killing vector. It must be noted that the charge was derived by using equation
of motion and so the analysis
is {\it on-shell}. Here we will present a completely {\it off-shell} analysis.

So far no attempts have been done to quantify the degrees of
freedom which are responsible to such universality. This issue we
will discuss in the context of Virasoro algebra and Cardy formula.
The method was first introduced in the context of gravity by Brown
and Hannueax \cite{Brown} and later developed by Carlip
\cite{Carlip1}. Although the method has been followed up in
several gravity theories \cite{Carlip2} -- \cite{Bibhas}, no
discussion exists in the presence of the one loop effective action
which leads to the trace anomaly. Here we will fill up this blank.
Our analysis will be followed from a recent work of one of the
authors \cite{Bibhas}. The key features of the
calculation are:\\
(i) The  whole analysis will be {\it off-shell};, i.e. no equation of motion will be used.\\
(ii) The derivation of the current corresponding to the total action (i.e. Einstein-Hilbert part plus
the one loop effective action) is {\it off-shell} and it is {\it off-shell} conserved.\\
These are essential since, as we have argued earlier, the
notion of entropy has general sense beyond the black hole horizon.

Let us first summarize the methodology. The main step is to 
define a bracket among the charges. In a previous
work of one of the authors \cite{Bibhas}, an off-shell definition 
of the bracket has been given for any general
covariant Lagrangian in terms of the arbitrary diffeomorphism vector and Noether current. Here we will use
this definition where explicit form of the current will be taken for the present theory. To evaluate it the
diffeomorphism vectors will be chosen using Carlip's formalism 
\cite{Carlip1}. This essentially tells that the vectors
 are chosen in such a way that they are asymptotically Killing vector near the horizon so that the
asymptotic killing horizon remains invariant.
It turns out that the Fourier modes of the bracket is similar to the standard form of the Virasoro algebra
which has central extension. Then the central charge and the 
zero mode eigenvalue are automatically identified.
Substituting these in Cardy formula \cite{Cardy} we obtain the expression for entropy. We will show that
{\it the entropy
will incorporate no correction for a classical background 
to the usual Bekenstein-Hawking expression}. Physically this analysis tells that some of the
degrees of freedom (DOF), which were originally gauge DOF, raise to true DOF which leads to entropy. Since
this is happening due to imposition of particular condition on the diffeomorphisms, the DOF responsible for
the entropy are observer dependent. This has been 
elaborated earlier in more details in \cite{Majhi,Majhi:2013jpk}.

The organization of our paper is as follows.
In section \ref{nogen} we present an analysis for a general covariant Lagrangian. First
the form of the Noether charge for a general covariant
Lagrangian is given.
Next we give the expressions for the Fourier modes of the charge and the central term corresponding
to the diffeomorphisms
which keep the Killing horizon structure invariant and are asymptotically Killing vector near the horizon.
Section \ref{noanom} is devoted to calculate these 
quantities explicitly for the anomalous effective action.
We then find the entropy using Cardy formula.
Finally, we conclude in section \ref{conclu}. For completeness, an appendix has been given at the 
end of the paper.
\section{Virasoro Algebra and Central Term: A General Approach}\label{nogen}

The recent progress in black hole thermodynamics shows that the
microscopic features may not be too sensitive to the details of
quantum gravity as the derivations for black hole temperature and
entropy only uses semiclassical gravity. In order to achieve this
it was first pointed out by Strominger that quantum state
calculation should rely on symmetry \cite{Strominger}. In an
earlier work, Brown and Henneaux \cite{Brown} showed that $(2+1)$
dimensional gravity has a asymptotic symmetry consisting of
Virasoro algebra with central extension, implying that any
microscopic quantum theory should be a conformal field theory.
Strominger noticed that use of the above central charge in the
Cardy formula \cite{Cardy} leads to the usual Bekenstein-Hawking
entropy. This implies that the asymptotic symmetries might shed
some light to the calculation of the density of states leading to
the statistical description of entropy. The limitations of the
earlier works are that they use the asymptotic AdS symmetry
which is insensitive to the structure of interior spacetime; i.e.
the details of the horizon. Moreover the method is confined to
($2+1$)-dimensions. Later on, Carlip introduced an approach which
relies on the Killing horizon structure preserving diffeomorphisms
and is independent of dimension of the spacetime \cite{Carlip1}.
So it reflects the information of the horizon which is important
because the entropy depends on the horizon structure. In this
paper, we shall adopt this approach.

In this section, we shall present the general expressions for the 
Fourier modes of Noether charge $Q$ and the central term
corresponding to a generally covariant Lagrangian. This method 
not only applies to Einstein gravity but also to any theories
of gravity whose Lagrangian is a function of metric, 
curvature tensor and some other external fields, like scalar field. 
The technique we shall follow from \cite{Carlip1,Bibhas}.

   Let us first introduce the general expressions for the Noether 
conserved current $J^a$ and the charge $Q$ corresponding to a 
generally covariant Lagrangian. The explicit variation of this general Lagrangian 
for the metric variation leads to the following form,
\begin{equation}\label{vir07}
\delta \left(L_{grav}\sqrt{-g}\right)=\sqrt{-g}
\left\lbrace E_{ab}\delta g^{ab}+\nabla _{a}\left(\delta v^{a} \right) \right\rbrace
\end{equation}
where $E_{ab} = 0$ leads to equation of motion and $\delta v^{a}$ is the surface term.
Now if the variation is given by the Lie derivative due to the coordinate transformation
$x^{a}\rightarrow x^{a}+\xi ^{a}$, then 
$\delta g_{ab}\equiv\pounds_{\xi}g_{ab}=\nabla _{a}\xi _{b}+\nabla _{b}\xi _{a}$.
Therefore, using the generalized Bianchi identity 
$\nabla _{a}E^{ab}=0$ \cite{pad11},
the first term on the right hand side of 
Eq. (\ref{vir07}) can be casted to a total derivative form 
$-2\sqrt{-g}\nabla_a(E^a_b\xi^b)$.
On the other hand, since $L_{grav}\sqrt{-g}$ is a scalar density, 
the Lie variation is given by $\sqrt{-g}\nabla_a(L_{grav}\xi^a)$.
Using all these in Equation (\ref{vir07}) we obtain a conservation 
relation given by $\nabla _{a}J^{a}=0$, where
\begin{equation}
J^{a}= \left(L_{grav}\xi ^{a}-\pounds_{\xi}v^{a}+2E^{ab}\xi _{b}\right)~.
\label{current}
\end{equation}
Here $\pounds_{\xi}v^{a}$ represents boundary term arising 
from the Lie variation of the metric as stated earlier.
The current $J_{a}$ is called Noether current.
Since the current is covariantly conserved, 
it can be expressed as the covariant derivative of an antisymmetric tensor:
$J^{a}=\nabla _{b}J^{ab}$, where $J^{ab}$ is known as the Noether potential.
For the general class of covariant gravitational theories, substituting the respective values in (\ref{current}) one can obtain the explicit expression.  
This leads to the current and potential as (See, Page $394$ of \cite{padbook}, for details):
\begin{eqnarray}J^{a}&=&\frac{1}{8\pi G}P^{abcd}\nabla _{b}\nabla _{c}\xi _{d}-\frac{1}{8\pi G}\nabla _{b}
\left(P^{adbc}+P^{acbd}\right)\nabla _{c}\xi _{d}-\frac{1}{4\pi G}\xi _{d}\nabla _{b}\nabla _{c}P^{abcd}
\label{vir08}
\\
J^{ab}&=&\frac{1}{8\pi G}P^{abcd}\nabla _{c}\xi _{d}-\frac{1}{4\pi G}\left(\nabla _{c}P^{abcd} \right)\xi _{d}
\label{vir09}
\end{eqnarray}
where four index tensor $P^{abcd}$ is defined as $P^{abcd}=\partial L_{grav}/\partial R_{abcd}$.
This tensor has identical symmetry properties as that of the curvature tensor $R^{abcd}$ i.e. antisymmetric
in interchange of $a,b$ and $c,d$, symmetric in interchange of pairs 
$(a,b)$ and $(c,d)$ along with $P^{a(bcd)}=0$. 
The corresponding Noether charge is defined as:
\begin{equation}\label{vir14}
Q[\xi]=\frac{1}{2}\int d\Sigma _{ab}\sqrt{h}J^{ab}
\end{equation}
where $d\Sigma _{ab}$ corresponds to the surface element. For instance, using 
Eq. (\ref{vir09}) one can show that the explicit expression for Noether
potential corresponding to the Einstein gravity appears to be,
$J^{ab}=(1/16\pi G) \left(\nabla ^{a}\xi ^{b}-\nabla ^{b}\xi ^{a}\right)$.
Now using this in Eq. (\ref{vir14}) for a timelike Killing vector field $\xi ^{a} = \chi^a$,
Noether charge $Q$, calculated on the horizon, can be shown to be 
related to the horizon entropy. More specifically, $Q$ multiplied by the
periodicity of the Euclidean time leads to Bekenstein-Hawking entropy expression: $(2\pi/\kappa)Q=A_h/4G$,
with $\kappa$ being the surface gravity. The above procedure has 
also been extended for Lanczos-Lovelock models to get black hole
entropy as Noether charge \cite{Bibhas}. 
However note that in all these models we have $\nabla _{a}P^{abcd}=0$. 
Having set the stage we shall now deal with the general expressions (\ref{vir08}) and (\ref{vir09})
to consider the situation where $\nabla_a P^{abcd}\neq 0$.

Next we give the expressions for the Fourier modes of the charge and the central 
extension. Applying the method of Carlip \cite{Carlip1} for stretched 
horizon scenario and using the definition of bracket among the 
charges, given in \cite{Bibhas}, 
we find the Fourier modes of the charge as,
\begin{eqnarray}
Q_{m}=-\frac{1}{32\pi G}\int \sqrt{h}d^{d-2}XP^{abcd}\mu _{ab}\mu _{cd}\left[2\kappa T_{m}-\frac{1}{\kappa}D^{2}T_{m}\right]-
\frac{1}{8\pi G}\int \sqrt{h}d^{d-2}X \mu _{ab}\chi _{d}T_{m}\left(\nabla _{c}P^{abcd} \right)
\label{FourierQ}
\end{eqnarray}
while the Fourier modes for the central term comes out to be:
\begin{eqnarray}
K[\xi _{m},\xi _{n}]&=&-\frac{1}{32\pi G}\int \sqrt{h}d^{d-2}X P^{abcd}\mu _{ab}\mu _{cd}\frac{1}{\kappa}
DT_{m}D^{2}T_{n}
-\frac{1}{8\pi G}\int \sqrt{h} d^{d-2}X \frac{\mid \chi \mid}{\rho}\frac{\chi _{b}\chi _{c}\rho _{d}}{\kappa \chi ^{2}}
T_{n}D^{2}T_{m}\nabla _{f}P^{bcfd}
\nonumber
\\
&-&\frac{1}{8\pi G}\int d^{d-2}X\sqrt{h}\frac{\mid \chi \mid}{\kappa ^{2}\chi ^{2}\rho}\chi _{b}\chi _{c}\rho _{d}
DT_{n}D^{2}T_{m}\nabla _{f}P^{bdfc}
\nonumber
\\
&+&\frac{1}{4\pi G}\int d^{d-2}X \sqrt{h} \frac{\mid \chi \mid}{\kappa \rho ^{3}}
\left(\chi ^{2}\rho _{b}\rho _{d}-\rho ^{2}\chi _{b}\chi _{d}\right)T_{n}DT_{m}\nabla _{f}\nabla _{c}P^{bfcd}
\nonumber
\\
&-&\frac{1}{4\pi G}\int \sqrt{h} d^{d-2}X \frac{\vert \chi \vert}{\rho \chi ^{2}}\chi _{a}\rho _{b}\chi _{d}
T_{m}DT_{n}\nabla _{c}P^{abcd}
\nonumber
\\
&-&\frac{1}{4\pi G}\int \sqrt{h} d^{d-2}X \frac{\vert \chi \vert}{\rho \chi ^{2}}
\left(\frac{\chi ^{2}}{\kappa \rho ^{2}}\right)^{2}DT_{m}D^{2}T_{n}\chi _{a}\chi _{b}\rho _{d}\nabla _{c}P^{abcd}-\left(m\leftrightarrow n\right)~.
\label{car05}
\end{eqnarray}
The detailed derivation of the above quantities can be followed 
from \cite{Bibhas}. But for the completeness and clarity, we have 
shown this in the appendix \ref{appA}.
We will use these expressions to identify the zero mode
eigenvalue and the central charge to calculate the entropy.
To evaluate the above, the explicit expression for $T_m$ is needed. 
This will be chosen subject to the 
condition that the Fourier modes for the diffeomorphism vector
$\xi^a$ obey the following subalgebra isomorphic to Diff $S^{1}$,
\begin{equation}\label{car01}
i\left\lbrace \xi _{m},\xi _{n}\right\rbrace ^{a}=(m-n)\xi _{m+n}^{a}
\end{equation}
with $\left\lbrace~,~\right\rbrace$ being the Lie bracket. 
Keeping this condition in mind, the form for $T_{m}$ can be taken as,
\begin{equation}\label{car06}
T_{m}=\frac{1}{\alpha} {\textrm{exp}}[im (\alpha t+g(x)+p.x_{\bot})]
\end{equation}
where $\alpha$ being a constant and $g(x)$ is a function which is regular at the Killing Horizon.
Here $p$ is an integer and $x_{\bot}$ is the ($d-2$) 
dimensional transverse co-ordinates with $t-x$ plane defining
the null surface. The explicit expression Equation 
(\ref{car06}) for $T_m$ will be needed later to evaluate the charge and the central term.
However, as we will see later, this will involve the choice for
the parameter $\alpha$. To obtain the correct expression for the
entropy one has to consider the periodicity of the Euclidean time
$2\pi /\kappa$. Then for periodicity in time
coordinate of Eq. (\ref{car06}), one must have $\alpha =\kappa$. This will be needed at
the end of the paper. For Lanczos-Lovelock gravity $\nabla
_{c}P^{abcd}$ vanishes and so the above expressions reduce to
those obtained earlier in \cite{Bibhas}. Here the results are much
more general and can be applied for the covariant gravity theory
for which the covariant derivative of $P^{abcd}$ does not vanish.
Moreover, the expressions are {\it off-shell} and free of any
ambiguity, as mentioned earlier.

Having obtained these general results we shall now apply these
results for anomaly induced effective action arising from the
fluctuations in the matter fields and background metric.  The
purpose of the next section is to calculate each terms of Eq.
(\ref{FourierQ}) and Eq. (\ref{car05}) for the anomalous effective
action and use them in the Cardy formula to derive the corrections
to the usual form of the entropy. The analysis will be done for
the case where the four dimensional trace anomaly appears. The
basic idea is that trace anomaly, being independent of
renormalization schemes and quantum states, it has some effects on
quantum correction to black hole area law. Furthermore, the
explicit expression for the anomaly can be derived from the
anomaly-induced effective action. Hence it would be interesting to
find the Noether charge and then use it to find the entropy.
Actually, to use Eq. (\ref{FourierQ}) and Eq. (\ref{car05}), we
need to find the explicit expression for $P^{abcd}$ for the action
on which we are interested. Then substitution of it will lead to
the final result.

\section{Anomalous effective action and entropy}\label{noanom}

In this section we shall briefly review the anomalous 
effective action for the fluctuations of the fields and consequently
the Noether charges are obtained thereof.
The classical General Relativity does not take 
into account the microscopic effects of quantum 
matter on any scale. In order to discuss
the wave like nature of a particle or phase correlation 
we need at least a semi classical treatment of effective stress energy tensor
appearing in Einstein's equation. The one-loop effective 
action could be determined by the method called Effective Field Theory (EFT).
It can be shown that non-local macroscopic coherence 
effects are actually self contained in low energy 
EFT provided the one-loop trace
anomaly for massless fields are included in Einstein 
Theory \cite{Mottola}. The low energy EFT of gravity 
contains an expansion in
derivatives of local term, while the higher order terms 
are suppressed by inverse ultraviolet cutoff scale $M$. 
Also the theory is not
renormalizable, however this is not sensitive to all 
the microscopic details due to decoupling of short distance degrees of freedom
\cite{Buchbinder}.
When we broke a classical symmetry by quantum trace 
anomaly, the decoupling of short and long distance physical situations using
standard technique fails. Thus an anomaly can have significant effect 
on low energy EFT. The necessity of trace anomaly for low energy
EFT of gravity can be understood from the behavior of different 
terms in effective gravitational action under global Weyl rescaling.
Thus addition of anomaly terms to low energy effective gravity 
action is consistent with both quantum theory and Equivalence principle
\cite{Fabbri}.

This search for quantum corrections to black hole entropy 
takes place in the context of quantum field theory in curved spacetime
\cite{Birrell1}. The backreaction of quantum fields on 
curved spacetime is determined by the semiclassical Einstein equation
\begin{equation}\label{not1}
G_{ab}+\Lambda g_{ab}=8\pi \langle T_{ab} \rangle
\end{equation}
such that the quantum fields affect the curved background 
through the expectation value of the energy-momentum tensor.
This expectation value can be obtained using one loop 
correction due to the fluctuations in the quantum fields to the classical
action and then taking the variation with respect to 
the metric. The trace of it is then given by (\cite{Balbinot}, \cite{Mottola}),
\begin{equation}\label{not02}
\langle T \rangle \equiv -\frac{2}{\sqrt{-g}}g^{ab}\frac{\delta S_{eff}}{\delta g^{ab}}=-\frac{2}{\sqrt{-g}}g^{ab}
\frac{\delta S_{anom}}{\delta g^{ab}}~.
\end{equation}
In four dimension there exists three geometric contribution 
to anomaly induced action, namely the Euler density (called Type-A),
the Weyl tensor squared (called Type-B) and a $\square R$ 
term that comes from variation of $R^{2}$. Here we will pay attention only
to type-A anomaly, since it is interesting in various ways 
for the study of black holes \cite{Fabbri} and also it is simple compared to other types.
This type of anomaly leads to trace anomaly:
\begin{equation}
\langle T \rangle = -\frac{a}{16\pi^2 G} E_4~,
\label{trace}
\end{equation}
where $E_4 = R_{abcd}R^{abcd}-4R_{ab}R^{ab}+R^{2}$. 
The corresponding low energy effective action is given by,
\begin{equation}\label{not03}
S_{eff}=-\frac{1}{16\pi G}\int d^{4}x \sqrt{-g} R +S_{anom}
\end{equation}
where the anomalous part of the action is of the form $S_{anom} = S_0+S_1+S_2+S_3$ with \cite{Aros},
\begin{eqnarray}\label{not05a}
&&S_{0}=-\frac{a}{32 \pi ^{2}G} \int d^{4}x \sqrt{-g} \left\lbrace -(\square \phi)^{2}\right\rbrace~;
\\
\label{not05}
&&S_{1}=-\frac{a}{16 \pi ^{2}G} \int d^{4}x \sqrt{-g}\left(R^{ab}-\frac{1}{3}Rg^{ab}\right)\nabla _{a}\phi \nabla _{b}\phi~;
\\
\label{not06}
&&S_{2}=\frac{a}{48 \pi ^{2}G} \int d^{4}x \sqrt{-g}\phi \square R~;
\\
\label{not07}
&&S_{3}=-\frac{a}{32 \pi ^{2}G} \int d^{4}x \sqrt{-g} E_{4}\phi~.
\end{eqnarray}
The scalar field $\phi$ satisfies the following equation of motion:
\begin{equation}
\left[\Box ^{2}+2\nabla _{\mu}\left( R^{\mu \nu}-\frac{1}{3}g^{\mu \nu}R \right) \nabla _{\nu} \right]\phi
=\frac{1}{2}\left(E_{4}-\frac{2}{3}\Box R\right)~.
\label{boxphi}
\end{equation}
Note that $S_{0}$ does not contain any curvature part 
and so for that part derivative with respect to $R_{abcd}$ vanishes. Thus there is
no Noether charge corresponding to this part. The standard way to get trace anomaly,  as first pointed
out by Polyakov \cite{Polyakov}, is to find the conformal primitive. 
In four dimensions we need two generalizations, firstly
the conformal properties of Gaussian curvature where D'Alembertian 
are determined by the Q-curvature defined
as $E_{4}-\frac{2}{3}\square R$ \cite{Branson} and secondly the Paneitz operator
\cite{Paneitz} should be introduced (see also \cite{Regge} and \cite{Katz}).

The entropy tensors can be obtained from the above actions 
by the usual prescription. These are given by
\begin{eqnarray}\label{not08}
&&P^{abcd}_{1}=-\frac{a}{32\pi ^{2}G}\left[\left(g^{ac}\nabla ^{b}\phi \nabla ^{d}\phi -
g^{ad}\nabla ^{b}\phi \nabla ^{c}\phi \right)-\frac{1}{3}\left(g^{ac}g^{bd}-g^{ad}g^{bc}\right) \right]
\\
\label{not09}
&&P^{abcd}_{2}=-\frac{a}{96\pi ^{2}G}\square \phi \left(g^{ac}g^{bd}-g^{ad}g^{bc}\right)
\\
\label{not10}
&&P^{abcd}_{3}=-\frac{a}{32\pi ^{2}G}\phi \left[R^{abcd}-4\left(g^{ac}R^{bd}-g^{ad}R^{bc}\right)+R\left(g^{ac}g^{bd}-g^{ad}g^{bc}\right) \right]
\end{eqnarray}
Along with the usual Noether charge due to Einstein-Hilbert action
we have three extra charges which are obtained by substitution of
the respective entropy tensors in Eq. (\ref{vir09}) and using the
definition Eq. (\ref{vir14}) for the charge . These turn out to
be,
\begin{eqnarray}
\label{QEH}
&&Q_{EH}=\frac{1}{16\pi G}\int d\Sigma _{ab}\nabla ^{a}\xi ^{b}
\\
\label{03a}
&&Q^{(1)}=-\frac{a}{16 \pi ^{2}G}\int d\Sigma _{cd}\left[\nabla ^{c}\xi^{p}\nabla ^{d}\phi \nabla _{p}\phi -
\frac{1}{3} (\nabla \phi)^{2}\nabla ^{c}\xi ^{d}\right]+{\textrm{terms containing}}~(\nabla _{c}P^{abcd})
\\
\label{04}
&&Q^{(2)}=\frac{a}{48\pi ^{2}G}\int d\Sigma _{cd}\nabla ^{c}\xi ^{d}\square \phi + {\textrm{terms containing}}~(\nabla _{c}P^{abcd})
\\
&&Q^{(3)}=-\frac{a}{32 \pi ^{2}G} \int d\Sigma _{ab}\left[R^{abcd}\nabla _{c}\xi _{d}+4\left(\nabla ^{a}\xi ^{d}R^{b}_{d}-
\nabla ^{d}\xi ^{a}R^{b}_{d}\right)+R\nabla ^{a}\xi ^{b}\right]
\nonumber
\\
&&+ {\textrm{terms containing}}~(\nabla _{c}P^{abcd})~.
\label{05}
\end{eqnarray}
Note that these are identical to those obtained in
\cite{Aros} except the terms containing the covariant derivative
of $P^{abcd}$. However, it can be shown that these terms behave as
${\mathcal{O}}(\chi ^{2})$ near the horizon for $\xi^a$, chosen to
be a timelike Killing vector $\chi^a$. For instance, let us
consider $P_2^{abcd}$, presented in Eq. (\ref{not09}). Then
$\nabla _{c}P^{abcd}=-(a/96\pi
^{2}G)\left(g^{ac}g^{bd}-g^{ad}g^{bc}\right)\nabla _{c}\Box \phi$
as covariant derivative of metric tensor vanishes. Now to evaluate
this term explicitly, for simplicity we take a general spherically
symmteric metric of the form $ds^{2}=-f(r)dt^{2}+dr^2/f(r) + r^2
d\Omega^2$. Then one can easily show that the relevant quantity
$\mu _{ab}\chi _{d}\nabla _{c}P^{abcd}$ is $\mathcal{O}(f)$ since
the timelike Killing vector in this case is given by $\chi
^{a}=(1,0,0,0)$. Similar conclusions hold for other parts also.
Hence the  terms, containing the covariant derivative of entropy
tensor, in the above equations are all of $\mathcal{O}(\chi ^{2})$
as $\chi^2 = -f$. Wald \cite{Wald} has first introduced the idea
that entropy could be calculated from the Noether charge by
choosing the diffeomorphism vector $\xi^a$ as the timelike Killing
vector $\chi^a$. In the original formulation we have two spacetime
boundaries, the asymptotic region and the horizon. The Noether
charges on the asymptotic region determines mass and angular
momentum, while that on the horizon determines the entropy. The
boundary conditions are very important for calculation of Noether
charge for they can get modified by modification of the boundary
terms. Remember that to determine entropy, we need to calculate
the charges (\ref{QEH}), (\ref{03a}), (\ref{04}) and (\ref{05}) on
the horizon defined by the relation $g_{ab}\chi^a\chi^b =0$. So
the ${\mathcal{O}}(\chi^2)$ terms do not contribute and hence one
can drop those terms.

Following the similar arguments as above, it is also 
possible to show that the terms containing the
covariant derivative of $P^{abcd}$ in Eq. (\ref{FourierQ}) 
and Eq. (\ref{car05}) are of the order $\chi^2$. So they will
not contribute near the horizon. Taking into account 
this fact and following the identical steps, employed in \cite{Bibhas}, 
we obtain from Eq. (\ref{FourierQ}) and Eq. (\ref{car05}) as,
\begin{eqnarray}
&&Q_m =\frac{\hat{A}}{8\pi G}\frac{\kappa}{\alpha}\delta _{m,0}
\label{Qm}
\\
&&K[\xi _{m},\xi _{n}]=-im^{3}\left[\frac{\hat{A}}{8\pi G}\frac{\alpha}{\kappa}\right]\delta _{m+n,0}
\label{cent}
\end{eqnarray}
where we have introduced the Wald entropy function quantity,
\begin{equation}\label{car07}
{\hat{A}}=-\frac{1}{2}\int \sqrt{h}d^{d-2}X P^{abcd}\mu _{ab}\mu _{cd}
\end{equation}
which leads to the horizon area in GR. To achieve Eq. (\ref{Qm}) and Eq. (\ref{cent}), 
Eq. (\ref{car06}) has been used and then the integration is done over the transverse 
coordinates. Collecting all these it is possible to find the following form,
\begin{equation}
i[Q_{m},Q_{n}]=(m-n)Q_{m+n}+\frac{C}{12}m^{3}\delta _{m+n,0}
\end{equation}
which is in the form of Virasoro algebra with central extension. $C$ is known as the central charge and is given by
\begin{equation}\label{genvir10}
\frac{C}{12}=\frac{\hat{A}}{8\pi G}\frac{\alpha}{\kappa}~.
\end{equation}
The zero mode eigenvalue from Equation (\ref{Qm}) turns out to be,
\begin{equation}\label{genvir11}
Q_{0}=\frac{\hat{A}}{8\pi G}\frac{\kappa}{\alpha}
\end{equation}
Therefore, we obtain the central charges and zero mode eigenvalues
from Eq. (\ref{genvir10}) and Eq. (\ref{genvir11}) for each term
of the action as,
\begin{eqnarray}
\label{28}
&&\frac{C^{EH}}{12}=\frac{A_{h}}{8\pi G}\frac{\alpha}{\kappa};~~~~~~~~~~~~~~~~~~~~~~~~~
Q^{EH}_{0}=\frac{A_{h}}{8\pi G}\frac{\kappa}{\alpha}.
\\
&&\frac{C^{(2)}}{12}=\frac{a}{12\pi ^{2} G}r_{h}\frac{\alpha}{\kappa}\left(\frac{d\phi}{dr}\right)_{r_{h}};
~~~~~~~ Q^{(2)}_{0}=\frac{a}{12\pi ^{2} G}r_{h}\frac{\kappa}{\alpha}\left(\frac{d\phi}{dr}\right)_{r_{h}}.
\\
&&\frac{C^{(3)}}{12}=-\frac{a}{2\pi G}\frac{\alpha}{\kappa}\phi _{r_{h}} \chi ;
~~~~~~~~~~~~~~~Q^{(3)}_{0}=-\frac{a}{2\pi G}\frac{\kappa}{\alpha}\phi _{r_{h}} \chi.
\end{eqnarray}
where the respective terms due to $S_{1}$ vanishes. To obtain the
above expressions the explicit forms of $P^{abcd}$, given by
(\ref{not08}), (\ref{not09}) and (\ref{not10}), have been used in
(\ref{car07}). $\chi$ is the two dimensional Euler characteristics
of the horizon and $\phi_{r_h}$ is the value of $\phi$ at the
horizon. Also in the above expression $A_{h}$ is the value of the
quantity defined in equation (\ref{car07}) with
$P^{abcd}=\frac{1}{2}\left(g^{ac}g^{bd}-g^{ad}g^{bc}\right)$.
Now substituting all these in the standard Cardy formula \cite{Cardy},
\begin{equation}\label{29}
S=2\pi \sqrt{\frac{C\Delta}{6}}; ~~~~~~\Delta \equiv Q_{0}-\frac{C}{24}
\end{equation}
we find the expressions for entropy:
\begin{eqnarray}
\label{30}
&& S^{EH}=\frac{A_{h}}{4G}
\\
&&S^{(2)}=\frac{a}{6\pi G}r_{h}\left(\frac{d\phi}{dr}\right)_{r_{h}}
\\
&&S^{(3)}=-\frac{a}{G}\phi _{r_{h}} \chi.
\end{eqnarray}
where we have used the identification $\alpha =\kappa$. The reason
has been explained earlier below Eq. (\ref{car05}). This can also
be followed from \cite{Bibhas}. Hence the total expression for
entropy corresponding to the anomaly induced action would be given
by,
\begin{equation}\label{31}
S_{tot}=\frac{A_{h}}{4G}+\frac{a}{6\pi G}r_{h}\left(\frac{d\phi}{dr}\right)_{r_{h}}-\frac{a}{G}\phi _{r_{h}} \chi
\end{equation}
The same expression was also obtained in \cite{Aros} by evaluating
the Noether charge for a Killing vector on the horizon. The charge
obtained by the authors of \cite{Aros} is {\it on-shell}. Here we
derived it from the Virasoro algebra technique. The analysis,
adopted here, is totally {\it off-shell}.

The remarkable fact about the above expression is, 
in contrary to \cite{Aros}, that the last two terms 
yield an additive constant to the entropy which can be neglected. This can be shown as follows.
For simplicity, let us consider the classical background as the
Schwarzschild metric. To obtain the explicit expression for $\phi$
in terms of metric coefficients, we need to solve Equation
(\ref{boxphi}). The solution will be found out by imposing certain
boundary conditions on the scalar field $\phi$. Since our only
concern is around the horizon we would require the finiteness of
$\phi$ and its first derivative at the horizon. Such conditions
lead the following solution \cite{Mottola}:
\begin{eqnarray}
\frac{d\phi}{dr}=-\frac{4Gm}{3r\left(r-2Gm\right)}\ln \frac{r}{2Gm}-\frac{1}{2Gm}-\frac{2}{r}~,
\label{schderphi}
\end{eqnarray}
where ${\textrm{dilog}}(x)=\int _{1}^{x}\frac{ln~(t)}{1-t}dt$. Integrating the above we find,
\begin{equation}
\phi (r)=-\frac{r}{2Gm}-2\ln \frac{r}{2Gm} + \frac{1}{3}\left(\ln \frac{r}{2Gm} \right)^{2}+\frac{2}{3}{\textrm{dilog}}
\Big(\frac{r}{2Gm}\Big)~.
\label{schphi}
\end{equation}
Note an important point that any constant term can shift the value
of $\phi$ at horizon. But the respective contribution to the
Noether charge and hence entropy would then vanish since the
action reduces to a topological term, which is the integral of
Euler density. Now it is easy to see that the term 
$\phi _{r_{h}}$ leads to a constant contribution 
and so the last term does not affect the Bekenstein-Hawking entropy. 
However $\frac{d\phi}{dr}$ leads to a term which is proportional 
to inverse of square root of horizon area. This would get 
compensated by the $r_{h}$ term. Hence the final expression 
for entropy leads to usual expression of entropy.
Therefore, from the above discussion, one can conclude 
that the type-A trace anomaly does not lead to any correction to entropy.

\section{Conclusions}\label{conclu}
   The idea that horizon entropy can be
obtained from certain class of diffeomorphism generators is pioneered mainly by the works
of Brown, Henneaux \cite{Brown} and Carlip \cite{Carlip1}. However all these methods use on-shell
criteria i.e. equations of motion have been used at one stage or
another. Recently, one of authors of this paper has
derived horizon entropy using Virasoro algebra and Cardy formula
bypassing the use of equation of motion \cite{Bibhas}. In the present work, as well we 
have given an \textit{off-shell} description i.e. no
equation of motion has been used. This again illustrates that the
notion of entropy is beyond black hole horizon.
We have started from the expressions for the Noether current and potential 
for an arbitrary generally covariant Lagrangian which could be function of 
metric tensor, curvature tensor and scalar
combination of different fields. From these, using the {\it off-shell} definition of 
bracket among the charges \cite{Bibhas}, a general expression for the central term was derived. 
We then obtained
the Fourier modes of the charge and the central term.

 Next we have evaluated the general expressions explicitly for the anomaly induced action 
which produces the type-A trace anomaly. This action has a pure
geometric meaning except the Euler characteristics and couplings 
appears in it. For this particular action we have shown that
the terms containing derivatives of $P^{abcd}$ vanishes in the
near horizon limit and the most dominating term is the one that
appears in Lanczos-Lovelock models. It has been shown that 
the Fourier modes of the bracket is similar to the usual Virasoro 
algebra with central extension. Identifying the central charge and the 
zero mode eigenvalue and then using them in the Cardy formula we obtained the 
expression for the entropy. The result is identical to that 
obtained in \cite{Aros}. Considering the classical background 
as the Schwarzschild metric, it has been shown that 
the entropy of the black hole is $A_{h}/4G$, the usual result. 
Hence type-A trace anomaly does not lead to any correction to horizon entropy.

  So far, we have observed that there is no correction term 
in the entropy. Our present analysis was based on the 
type-A trace anomaly. As we mentioned earlier, there also 
exists other types of trace anomalies, like type-B etc. 
It would be interesting to study the effective actions 
corresponding to these anomalies in the present direction 
and investigate if they lead to any correction to entropy. 
This we leave for future.

\section*{Acknowledgements}

We thank T. Padmanabhan for his several useful comments on the first draft of our paper. 
The research of one of the authors (BRM) is supported by a Lady Davis Fellowship at Hebrew University,
by the I-CORE Program of the Planning and Budgeting Committee and the Israel Science Foundation 
(grant No. 1937/12), as well as by the Israel Science Foundation personal grant No. 24/12. 
The research of S.C is funded by SPM Fellowship from CSIR, Govt. of India.


\appendix

\section{Detailed calculation for the Fourier modes of  the charge and the central term}\label{appA}

In this appendix, using Eq. (\ref{vir08}) and Eq. (\ref{vir09}) 
we shall obtain the expressions (\ref{FourierQ}) and (\ref{car05}), 
the Fourier modes of the Noether 
charge and the central term.
To start with, we consider a $d$-dimensional spacetime manifold $M$ with boundary
given as $\partial M$, such that neighborhood of $\partial M$
admits Killing vector $\chi ^{a}$, satisfying $\chi
^{2}=g_{ab}\chi ^{a}\chi ^{b}=0$ over the boundary $\partial M$.
We shall work in a stretched horizon method as used
by Carlip \cite{Carlip1}. In this approach, all the calculations
will be done on a boundary $\chi ^{2}=\epsilon$ and at the end the
limit $\epsilon \rightarrow 0$ will be imposed to obtain our final
results. Near this stretched horizon one can define an orthogonal
vector $\rho^a$ to $\chi ^{a}$ such that,
\begin{equation}\label{vir01}
\nabla _{a}\chi ^{2}=-2\kappa \rho _{a}
\end{equation}
with $\kappa$ being the surface gravity at the horizon. 
Next choose the diffeomorphism vector as,
\begin{equation}\label{vir05}
\xi ^{a}=T\chi ^{a}+R\rho ^{a}
\end{equation}
where $R$ and $T$ are the arbitrary functions of all coordinates
of spacetime. This diffeomorphism could be interpreted as
deformations in the $r-t$ plane, playing a crucial role in the
Euclidean approach to black hole thermodynamics \cite{Banados}.
The condition $\chi ^{a}\chi ^{b}\delta g_{ab}/\chi ^{2}\rightarrow 0$, when 
imposed on $\chi ^{a}$ leads to a relation between them:
\begin{equation}\label{vir06}
R=\frac{\chi ^{2}}{\kappa \rho ^{2}}DT
\end{equation}
where $D\equiv \chi ^{a}\nabla _{a}$. The condition is chosen in such a way that the asymptotic Killing horizon structure remains invariant after the perturbation. The diffeomorphism formed by
Eq. (\ref{vir05}) and Eq. (\ref{vir06}) is said to form a closed
subalgebra provided $\rho ^{a}\nabla _{a}T=0$ is satisfied near
the horizon. In this setup, we can define the surface element on
the horizon as \cite{Bibhas}
\begin{equation}
 d\Sigma _{ab}=d^{d-2}X\mu _{ab}; \,\,\ \mu _{ab}=-\frac{\mid \chi \mid}{\rho \chi ^{2}}\left(\chi _{a}\rho _{b}-\chi _{b}\rho _{a}\right)~.
\label{surfaceelement}
\end{equation}
To evaluate the charge and the bracket we also need 
three identities which are valid upto ${\cal{O}}(\chi^2)$.
The detailed derivations are given in \cite{Bibhas} 
(see, Eq. (B37), Eq. (B43) and Eq. (B38)). The identities are
\begin{eqnarray}
&&\nabla _{a}\xi _{b}=\frac{\chi _{a}\chi _{b}}{\chi ^{2}}DT+\frac{\kappa}{\chi ^{2}}\left(\chi _{a}\rho _{b}-\chi _{b}\rho _{a}\right)T
-\frac{1}{\kappa \chi ^{2}}\chi _{a}\rho _{b}D^{2}T+R\nabla _{a}\rho _{b}
\label{identity}
\\
&&P^{abcd}\nabla _{c}\xi _{d}=P^{abcd}\left[\frac{2\kappa}{\chi ^{2}}T-\frac{1}{\kappa \chi ^{2}}D^{2}T\right]\chi _{c}\rho _{d}
\label{identity1}
\\
&&\nabla _{d}\nabla _{a}\xi _{b}=\frac{2\kappa}{\chi ^{4}}\chi _{a}\rho _{b}\chi _{d}DT-
\frac{1}{\kappa \chi ^{4}}\chi _{a}\rho _{b}\chi _{d}D^{3}T-\frac{1}{\chi ^{4}}\chi _{a}\chi _{b}\chi _{d}D^{2}T~.
\label{identity2}
\end{eqnarray}
We shall now use the above expressions to find the charge and the bracket.
For a covariant Lagrangian, we define the bracket among the charges as
\begin{equation}\label{vir12}
\left[Q_{1},Q_{2}\right]=\left(\delta _{\xi _{1}}Q[\xi _{2}]-\delta _{\xi _{2}}Q[\xi _{1}]\right)\equiv
\int \sqrt{h}d\Sigma _{ab}\left[\xi _{2}^{a}J_{1}^{b}-\xi _{1}^{a}J_{2}^{b}\right]
\end{equation}
where the notation $J_{1}^{b}=J^{b}[\xi _{1}]$ has been used. 
The above expression is
{\it off-shell} and free of any ambiguity. The only fact needed is
that, the Noether current can be expressed as the covariant
derivative of a antisymmetric tensor (see Ref. \cite{Bibhas} for more details). To evaluate it for the
present setup, we first calculate the current using Eq.
(\ref{vir08}). It can be shown that the current, presented in
equation (\ref{vir08}), takes the following form:
\begin{eqnarray}
J^{a}&=&\frac{1}{8\pi G} P^{abcd}\xi _{b}\xi _{c}\rho _{d}\frac{1}{\xi _{4}}\left(2\kappa DT-\frac{1}{\kappa}D^{3}T\right)
\nonumber
\\
&-&\frac{1}{8\pi G}\left[\frac{\kappa}{\chi ^{2}}\left(\chi _{c}\rho _{d}-\chi _{d}\rho _{c} \right)T +
\frac{\rho _{c}\rho _{d}}{\chi ^{2}}DT-\frac{\chi _{c}\rho _{d}}{\kappa \chi ^{2}}D^{2}T \right] \nabla _{b}\left(P^{adbc}+P^{acbd}\right)
\nonumber
\\
&-&\frac{1}{4 \pi G}\xi _{d}\nabla _{b}\nabla _{c}P^{abcd}~.
\label{vir10}
\end{eqnarray}
In the intermediate steps, the identities (\ref{identity}), (\ref{identity1}) 
and (\ref{identity2}) have been used.
Then using Eq. (\ref{vir05}) and Eq. (\ref{surfaceelement}) we obtain,
\begin{eqnarray}
d\Sigma _{ab}\xi _{2}^{a}J_{1}^{b}&=&d^{d-2}X\frac{1}{32\pi G}\frac{\rho}{\mid \chi \mid}P^{abcd}\mu _{ab}\mu _{cd}
\left(2\kappa DT_{1}-\frac{1}{\kappa}D^{3}T_{1}\right)T_{2}
\nonumber
\\
&+&\frac{1}{8\pi G}d^{d-2}X\left(\frac{\mid \chi \mid}{\rho \chi ^{2}} \right)
\left(\chi _{a}\rho _{b}-\chi _{b}\rho _{a}\right)\xi _{2}^{a}\Big[\frac{\kappa}{\chi ^{2}}\left(\chi _{c}\rho _{d}-\chi _{d}\rho _{c} \right)T_{1}
\nonumber
\\
&+&\frac{\rho _{c}\rho _{d}}{\chi ^{2}}DT_{1}-\frac{\chi _{c}\rho _{d}}{\kappa \chi ^{2}}D^{2}T_{1} \Big]\nabla _{f}
\left(P^{bdfc}+P^{bcfd} \right)
\nonumber
\\
&+&\frac{1}{4\pi G}d^{d-2}X\left(\frac{\chi}{\rho \chi ^{2}} \right)\left(\chi _{a}\rho _{b}-\chi _{b}\rho _{a}\right)
\xi _{2}^{a}\xi _{1d}\nabla _{f}\nabla _{c}P^{bfcd}
\label{vir11}
\end{eqnarray}
where the relation 
$P^{abcd}\mu _{ab}\mu _{cd}=\frac{4}{\rho ^{2}\chi ^{2}}P^{abcd}\rho _{a}\chi _{b}\rho _{c}\chi _{d}$ 
has been used.
Finally, substitution of Eq. (\ref{vir11}) in Eq. (\ref{vir12}) leads to,
\begin{eqnarray}
\left[Q_{1},Q_{2}\right] &=& \frac{1}{32\pi G}\int \sqrt{h}d^{d-2}XP^{abcd}\mu _{ab}\mu _{cd}
\left[\frac{1}{\kappa}T_{1}D^{3}T_{2}-2\kappa T_{1}DT_{2} \right]
\nonumber
\\
&-&\frac{1}{8\pi G}\int \sqrt{h} d^{d-2}X \frac{\mid \chi \mid}{\rho}\frac{\chi _{b}\chi _{c}\rho _{d}}{\kappa \chi ^{2}}
T_{2}D^{2}T_{1}\nabla _{f}P^{bcfd}
\nonumber
\\
&-&\frac{1}{8\pi G}\int d^{d-2}X\sqrt{h}\frac{\mid \chi \mid}{\kappa ^{2}\chi ^{2}\rho}\chi _{b}\chi _{c}\rho _{d}
DT_{2}D^{2}T_{1}\nabla _{f}P^{bdfc}
\nonumber
\\
&+&\frac{1}{4\pi G}\int d^{d-2}X \sqrt{h} \frac{\mid \chi \mid}{\kappa \rho ^{3}}
\left(\chi ^{2}\rho _{b}\rho _{d}-\rho ^{2}\chi _{b}\chi _{d}\right)T_{2}DT_{1}\nabla _{f}\nabla _{c}P^{bfcd}
\nonumber
\\
&-&\left(1\leftrightarrow 2\right)
\label{vir13}
\end{eqnarray}
Next we shall obtain the expression for the charge $Q[\xi]$ in the
near horizon limit. The charge is given by Eq. (\ref{vir14}). Use
of Eq. (\ref{vir05}) and Eq. (\ref{identity2}) in Eq.
(\ref{vir14}) leads to
\begin{equation}\label{vir15}
Q[\xi]=-\frac{1}{32\pi G}\int \sqrt{h}d^{d-2}XP^{abcd}\mu _{ab}\mu _{cd}\left[2\kappa T-\frac{1}{\kappa}D^{2}T\right]-
\frac{1}{8\pi G}\int \sqrt{h}d^{d-2}X \mu _{ab}\xi _{d}\left(\nabla _{c}P^{abcd} \right)~.
\end{equation}
Now the central term will be calculated. This is defined as
\begin{equation}\label{vir16}
K[\xi _{1},\xi _{2}]=[Q_{1},Q_{2}]-Q[\left\lbrace \xi _{1},\xi _{2} \right\rbrace]
\end{equation}
where the quantity $[Q_{1},Q_{2}]$ given in Eq. (\ref{vir13}) and
$Q[\left\lbrace \xi _{1},\xi _{2} \right\rbrace]$ will be
evaluated using (\ref{vir15}) in the following way. Using Eq.
(\ref{vir05}), the Lie bracket near horizon turns out to be,
\begin{equation}\label{vir17}
\left\lbrace \xi _{1},\xi _{2} \right\rbrace ^{a}=\left\lbrace T_{1},T_{2} \right\rbrace \chi ^{a}+
\left\lbrace R_{1},R_{2} \right\rbrace \rho ^{a}
\end{equation}
where we have $\left\lbrace T_{1},T_{2}
\right\rbrace=\left(T_{1}DT_{2}-T_{2}DT_{1}\right)$ and so on.
Substituting this in Eq. (\ref{vir15}) we find,
\begin{eqnarray}
Q[\left\lbrace \xi _{1},\xi _{2}\right\rbrace]&=&-\frac{1}{32\pi G}\int \sqrt{h}d^{d-2}XP^{abcd}\mu _{ab}\mu _{cd}\Big[2\kappa
T_{1}DT_{2}
-\frac{1}{\kappa}DT_{1}D^{2}T_{2}+T_{1}D^{3}T_{2} \Big]-\left(1\leftrightarrow 2\right)
\nonumber
\\
&-&\frac{1}{8\pi G}\int \sqrt{h}d^{d-2}X \mu _{ab}\left\lbrace \xi _{1},\xi _{2}\right\rbrace _{d}\left(\nabla _{c}P^{abcd} \right)
\label{vir18}
\end{eqnarray}
Finally, substitution of Eq. (\ref{vir13}) and Eq. (\ref{vir18})
in Eq. (\ref{vir16}) yields the form of the central term as,
\begin{eqnarray}
K[\xi _{1},\xi _{2}]&=&-\frac{1}{32\pi G}\int \sqrt{h}d^{d-2}X P^{abcd}\mu _{ab}\mu _{cd}
\frac{1}{\kappa}DT_{1}D^{2}T_{2}-\frac{1}{8\pi G}\int \sqrt{h} d^{d-2}X
\frac{\mid \chi \mid}{\rho}\frac{\chi _{b}\chi _{c}\rho _{d}}{\kappa \chi ^{2}}
T_{2}D^{2}T_{1}\nabla _{f}P^{bcfd}
\nonumber
\\
&-&\frac{1}{8\pi G}\int d^{d-2}X\sqrt{h}\frac{\mid \chi \mid}{\kappa ^{2}\chi ^{2}\rho}\chi _{b}\chi _{c}\rho _{d}
DT_{2}D^{2}T_{1}\nabla _{f}P^{bdfc}
\nonumber
\\
&+&\frac{1}{4\pi G}\int d^{d-2}X \sqrt{h} \frac{\mid \chi \mid}{\kappa \rho ^{3}}
\left(\chi ^{2}\rho _{b}\rho _{d}-\rho ^{2}\chi _{b}\chi _{d}\right)T_{2}DT_{1}\nabla _{f}\nabla _{c}P^{bfcd}
\nonumber
\\
&-&\frac{1}{4\pi G}\int \sqrt{h} d^{d-2}X \frac{\vert \chi \vert}{\rho \chi ^{2}}\Big[\chi _{a}\rho _{b}\chi _{d}T_{1}DT_{2}\nabla _{c}P^{abcd}
+\left(\frac{\chi ^{2}}{\kappa \rho ^{2}}\right)^{2}DT_{1}D^{2}T_{2}
\chi _{a}\chi _{b}\rho _{d}\nabla _{c}P^{abcd}\Big]
\nonumber
\\
&-&\left(1\leftrightarrow 2 \right)~.
\label{vir19}
\end{eqnarray}

Now the Fourier modes of the Noether charge 
and the central term will be evaluated.
For this we need to first give the Fourier modes of the
arbitrary function $T$.
The Fourier decomposition of the functions $T_{1}$ and $T_{2}$ are taken as,
\begin{equation}\label{car02}
T_{1}=\sum _{m}A_{m}T_{m};~~~~~~~~~~~~~~~T_{2}=\sum _{n}B_{n}T_{n}
\end{equation}
with reality criteria for $T_{1}$ and $T_{2}$ being given by, $A_{n}^{*}=A_{-n}$ and $B_{m}^{*}=B_{-m}$.
Substituting Eq. (\ref{car02}) in Eq. 
(\ref{vir15}) and using $Q[\xi] =\displaystyle\sum_m A_m Q_m$,
we find the Fourier modes of the charge as given in Eq. (\ref{FourierQ}).
Similarly, in order to obtain the same for the central term, we define
\begin{equation}\label{car04}
K[\xi _{1},\xi _{2}]=\sum _{m,n}C_{m,n}K[\xi _{m},\xi _{n}]
\end{equation}
with $C_{m,n}\equiv A_{m}B_{n}$ and so $C_{m,n}^{*}\equiv C_{-m,-n}$. 
This manipulation in turn leads to the Fourier decomposition of the 
central term as presented by Eq. (\ref{car05}).


\end{document}